# Spectroscopic evidence for electron-boson coupling in electron-doped Sr$_2$IrO$_4$

Yong Hu[1, #], Xiang Chen[2, #], S.-T. Peng[1], C. Lane[3], M. Matzelle[3], Z.-L. Sun[1], M. Hashimoto[4], D.-H. Lu[4], E. F. Schwier[5], M. Arita[5], T. Wu[1], R. S. Markiewicz[3], K. Shimada[5], X.-H. Chen[1], Z.-X. Shen[4,6], A. Bansil[3], S. D. Wilson[2] and J.-F. He[1,*]

[1]Hefei National Laboratory for Physical Sciences at the Microscale, Department of Physics and CAS Key Laboratory of Strongly-coupled Quantum Matter Physics, University of Science and Technology of China, Hefei, Anhui 230026, China

[2]Materials Department, University of California, Santa Barbara, California 93106, USA

[3]Department of Physics, Northeastern University, Boston, Massachusetts 02115, USA

[4]Stanford Synchrotron Radiation Lightsource and Stanford Institute for Materials and Energy Sciences, SLAC National Accelerator Laboratory, Menlo Park, California 94025, USA

[5]Hiroshima Synchrotron Radiation Center, Hiroshima University, Hiroshima 739-0046, Japan

[6]Geballe Laboratory for Advanced Materials, Departments of Physics and Applied Physics, Stanford University, Stanford, California 94305, USA

[#]These authors contributed equally to this work.

[*]To whom correspondence should be addressed: jfhe@ustc.edu.cn

**The pseudogap, d-wave superconductivity and electron-boson coupling are three intertwined key ingredients in the phase diagram of the cuprates. Sr$_2$IrO$_4$ is a 5d-electron counterpart of the cuprates in which both the pseudogap and a d-wave instability have been observed. Here, we report spectroscopic evidence for the presence of the third key player in electron-doped Sr$_2$IrO$_4$: electron-boson coupling. A kink in nodal dispersion is observed with an energy scale of ~50 meV. The strength of the kink changes with doping, but the energy scale remains the same. These results provide the first noncuprate platform for exploring the relationship between the pseudogap, d-wave instability and electron-boson coupling in doped Mott insulators.**

A central issue in cuprate high-temperature superconductors is to understand the presence of multiple orders in their phase diagrams and how these orders are driven by various interactions. The mysterious pseudogap phase is intertwined with d-wave superconductivity as either a competitor or a precursor [1-3]. Electron-boson coupling is another key player, which has been widely recognized as the pairing glue for realizing superconductivity [1, 4, 5]. Recent experimental evidence has further revealed a close connection between the electron-boson coupling and the pseudogap [6, 7] – thus forming an interacting loop between these three key players in the phase diagram. In $Bi_2Sr_2CaCu_2O_{8+\sigma}$, a specific positive-feedback loop has been proposed between the electron-boson (phonon) coupling and electronic orders in the pseudogap phase, which in turn could enhance the d-wave superconductivity [7]. In this sense, the coexistence of the pseudogap, d-wave superconductivity and electron-boson coupling, as well as the interactions between them, are of key importance in unraveling the exotic physics of the cuprates.

Ruddlesden-Popper strontium iridate $Sr_2IrO_4$ is a pseudospin-1/2 Mott insulator due to the cooperative action of spin-orbit coupling and on-site Coulomb interactions [8-10]. Theoretically, it has been described by the same minimal model as that for the cuprates [11, 12]. Therefore, it is natural to ask whether or not the exotic quantum phenomena observed in the cuprates can also be seen in this non-cuprate material. Much recent work has gradually unveiled structural, electronic and magnetic parallels between these two material classes [13-20]. In particular, the mysterious pseudogap as well as the d-wave instability have been observed in electron-doped $Sr_2IrO_4$ [14, 15, 21], mimicking the hallmarks of hole-doped cuprates [1,22-25].

In this paper, we report spectroscopic evidence for the presence of the third key player in $Sr_2IrO_4$: electron-boson coupling. A dispersion kink is observed at ~50 meV along the nodal direction, accompanied by a sharpening of the quasiparticle peak below the kink energy. This feature is similar to that observed in the cuprates and indicates the existence of electron-boson coupling. Doping dependent measurements show that the coupling strength changes with doping, but the mode energy stays the same within the error bars. These observations establish another significant similarity between $Sr_2IrO_4$ and cuprates. The coexistence of the pseudogap, d-wave instability and electron-boson coupling in electron-doped $Sr_2IrO_4$ also provides a new window to investigate the interactions between these exotic quantum phenomena in doped spin-orbit coupled Mott insulators.

Single crystals of $(Sr_{1-x}La_x)_2IrO_4$ were grown via a platinum (Pt) crucible-based flux growth method as described in an earlier study [19]. X-ray diffraction measurements were carried out at room temperature to exclude the presence of a possible $Sr_3Ir_2O_7$ phase. The La-doping levels were determined via energy-dispersive X-ray spectroscopy (EDS) measurements (see Supplemental Material, Fig. S1). Bulk La-doped metallic samples $(Sr_{1-x}La_x)_2IrO_4$ (x~0.04) were used as the starting point for *in situ* surface doping in order to avoid possible charging of the bulk material during angle-resolved photoemission (ARPES) measurements. Samples were cleaved at 30 K in ultrahigh vacuum. Continuous electron doping was realized by *in situ* potassium deposition. ARPES results were obtained at Beamline 5-4 of the Stanford Synchrotron Radiation Lightsource (SSRL) of SLAC National Accelerator Laboratory using 25 eV photons with a total energy resolution of ~7 meV and a base pressure of better than $3 \times 10^{-11}$ torr. The Fermi level was obtained by measuring polycrystalline Au in electrical contact with the sample. Some related preliminary tests were performed at Hiroshima Synchrotron Radiation Center (HSRC).

*Ab initio* calculations were carried out by using the pseudopotential projector-augmented wave method implemented in the Vienna ab initio simulation package (VASP) with an energy cutoff of 500 eV for the plane-wave basis set [26, 27]. Exchange-correlation effects were treated using the generalized gradient approximation, where a 12 x 12 x 3 Γ-centered k-point mesh was used to sample the Brillouin zone. The total energy was converged with a tolerance of $10^{-5}$ eV. Spin-orbit coupling effects were included self-consistently. We used the low-temperature *I4/mmm* crystal structure in accord with experimental observations.

The measured photoelectron intensity as a function of energy and momentum for an electron-doped $Sr_2IrO_4$ sample along the (0, 0) – (π, π) nodal direction is shown in Fig. 1(a) (also see Supplemental Material, Fig. S2). A low energy kink is seen in the main band near (π/2, π/2), marked by the black arrow (also see Supplemental Material, Fig. S3). Quantitative extraction of the dispersion is obtained by fitting the momentum distribution curves (MDCs). In Fig. 1(b), one can see a clear dispersion kink at ~50 meV below the Fermi level ($E_F$). We note that the folded band is well separated from the main band near the kink energy [Fig. 1(a)], thus it does not affect the identification of the kink in the main band. An energy scale in the band dispersion should also manifest itself as an energy feature in the electron self-energy. We follow the analysis in cuprates [28-30] to extract the effective real part of the electron self-energy (Re $\Sigma$) from dispersion by assuming a straight line as the featureless bare band. As shown in Fig. 1(c), a prominent peak appears at ~50 meV, confirming the energy scale of the dispersion kink. We note that the energy of the peak does not change with the selection of the empirical bare band (see Supplemental Material, Fig. S4 for details). This is similar to what is observed in the cuprates, and points to an intrinsic renormalization of the band dispersion.

A concomitant observation is the sharpening of the quasiparticle peak below the kink energy [~50 meV, see Fig. 1(a)]. For this purpose, the peak width of the energy distribution curves (EDCs) is plotted as a function of energy in Fig. 1(d). A drop in the width is seen below ~50 meV, indicating a reduction of the scattering rate (see Supplemental Material, Fig. S3). This scattering rate change can also be visualized by the MDC peak width [28]. In Fig. 1(e), the MDC peak width is shown as a function of energy, where a drop at ~50 meV is discernible. We note that the drop of the MDC peak width in electron-doped $Sr_2IrO_4$ is not as strong as that in hole-doped cuprates (e.g., $Bi_2Sr_2CaCu_2O_{8+\sigma}$ [28]), but similar to what is observed in electron-doped cuprates (e.g., $Nd_{2-x}Ce_xCuO_4$ [31]). This scattering rate reduction at ~50 meV represents a decrease in the imaginary part of the electron self-energy (Im $\Sigma$), which echoes the unveiled peak in Re $\Sigma$ at the same energy.

Doping evolution of the nodal dispersion was studied via *in situ* potassium deposition and the results are shown in Fig. 2(a-c). As established in earlier studies [14, 21] and confirmed by our experiments [Fig. 2(d-e)], electron doping can be effectively induced by potassium deposition on the $Sr_2IrO_4$ sample surface (also see Supplemental Material, Fig. S5). In order to quantify the doping dependence of the kink, MDC- derived dispersion of the main band is extracted at each doping level [Fig. 3(a-c)]. It is clear that the kink stays at the same binding energy as a function of doping (marked by the black arrow). However, the band renormalization caused by the kink seems to increase with increasing doping level. This is quantified in the extracted effective Re $\Sigma$ [Fig. 3(d-f)]. While the prominent peak stays at ~50 meV (Fig. 4), the overall magnitude of the self-energy continuously increases with electron doping, indicating a stronger deviation from the bare band. As demonstrated in cuprates, the strength of the band renormalization can also be quantified by taking the ratio between the "high-energy" velocity above the kink energy and the dressed velocity below the kink energy [30]. In order to parallel the methodology used in the cuprates, we define this ratio as ($\lambda'+1$), and plot $\lambda'$ as a function of doping (Fig. 4). It is clear that $\lambda'$ increases with electron doping, suggesting an enhanced renormalization of the band with doping. This trend is found to be monotonic up to the highest doping level we have achieved. We note that the kink feature is not well defined in the low doping regime due to lack of quasiparticles along the nodal direction [15].

We next turn to discuss the origin of the observed dispersion kink. The first thing to check is whether the observed kink can be simply attributed to a curved bare band dispersion. This is unlikely for the following reasons. First, subtracting a straight line from a curved dispersion can indeed produce an "effective Re Σ", but we will not expect an energy feature to appear (see Supplemental Material, Fig. S4). This is distinct from the experimental observation of a robust peak feature in the effective Re Σ, irrespective of the empirical bare band selected (either straight lines or the bare band from first-principles calculations, see Supplemental Material, Fig. S4). Second, even if one assumes an artificial bare-band dispersion with a feature and attributes the kink to such a bare band feature, then the binding energy of the kink should move with electron doping (because the chemical potential moves with doping), and the strength of the kink should stay the same. These expectations are, however, in sharp contrast with our experimental results. Third, any possible bare band feature cannot give rise to the observed scattering rate reduction below the kink energy. Another special property of $Sr_2IrO_4$ is the existence of octahedral rotation, which could give rise to the folded band [15]. However, the kink is clearly identified in the main band which makes the octahedral rotation irrelevant in this connection.

A more plausible explanation involves the presence of an energy scale from electron-boson coupling, similar to that reported in the cuprates [28-30, 32]. Here, the band dispersion is normalized by the electron-boson coupling and the kink marks the energy of the boson mode. This mode coupling naturally explains the observed peak in Re Σ and the drop in Im Σ (scattering rate). Then, the experimentally extracted λ' represents the effective coupling constant [29, 30] (although it is an overestimate of the real coupling constant λ, as pointed out in [30]). As for the origin of the boson, the first possibility is a phonon, since there are phonons of this energy scale in electron-doped $(Sr_{1-x}La_x)_2IrO_4$. In Raman scattering measurements, the most prominent phonon mode ($B_{2g}$) locates at ~50 meV [33], which is identical to the energy of the observed dispersion kink. Temperature dependent measurements indicate that the $B_{2g}$ mode persists to room temperature, which would also explain our observation of the dispersion kink at elevated temperatures (Supplemental Material, Fig. S6). However, this phonon scenario cannot naturally explain why the kink strength increases with electron doping. In the traditional picture, the strength of electron-phonon coupling should decrease with doping due to an enhanced screening, although there are examples of anti-screening, where the coupling strength for a particular phonon mode becomes stronger with doping [34]. Whether the anti-screening is at play in electron-doped $Sr_2IrO_4$ remains to be explored. The second possible candidate comes from magnetic excitations. We note that the energy of the magnetic excitation at Q=(π/2, π/2) is close to 50 meV in electron-doped $(Sr_{1-x}La_x)_2IrO_4$ [35, 36]. The wavevector Q=(π/2, π/2) might also correspond to a low energy scattering process between the nodal and antinodal regions. However, the 50 meV scale is only part of the dispersion of the magnon modes in $(Sr_{1-x}La_x)_2IrO_4$. It is distinct from the resonance mode in the cuprates that forms a peak in the local spin susceptibility [37, 38]. Moreover, the energy of the magnetic excitation at Q=(π/2, π/2) shows a moderate change as a function of doping, which is different from the doping independent energy scale marked by the dispersion kink in our experiment. Whether this discrepancy can be reconciled by the different absolute doping levels between our study and the RIXS studies on magnetic excitations [35, 36] is not clear. Another explanation may involve a possible fluctuation of the hidden order reported in $Sr_2IrO_4$ [39], with the observed energy scale in our experiments reflecting the frequency of the fluctuation.

Finally, we comment on the possible relationship between the electron-boson coupling, d-wave instability and the pseudogap in electron-doped $Sr_2IrO_4$. Different from the electron-boson coupling (e.g., electron-phonon coupling) in many weakly correlated materials which can be described by the textbook example, the electron-boson coupling in strongly correlated systems (e.g., cuprates) is much more complicated. In the cuprates, the electron-boson coupling has been proposed as the pairing glue for high $T_c$ d-wave superconductivity [1, 4, 5]. The same type of superconductivity has been theoretically predicted in electron-doped $Sr_2IrO_4$ [11, 12], and a d-wave instability has been experimentally observed in heavily electron-doped regime via surface potassium deposition [21]. It is interesting, therefore, to ask whether the electron-boson coupling in electron-doped $Sr_2IrO_4$, which has an energy scale similar to the cuprates, enhances the tendency toward d-wave instability [7], and whether this instability represents intrinsic superconductivity [21, 40]. In the cuprates, it has also been suggested that the electron-boson (phonon) coupling and the pseudogap related electronic correlations reinforce each other due to their synchronized response in the same doping regime [7]. In $Sr_2IrO_4$, the dispersion kink appears in a doping range close to the pseudogap region [14, 15]. Whether they develop in a synchronized fashion in a certain doping range is yet to be explored. Also, a variety of modes have been reported in the cuprates, which could play very different roles [41]. It would be interesting to investigate the possible existence of multiple mode-couplings and their momentum dependencies in $Sr_2IrO_4$ [42]. Despite the similarities to the cuprates, $Sr_2IrO_4$ presents unique features driven by the presence of strong spin-orbit coupling effects. For example, a spin-orbit-controlled metal-insulator transition was recently reported in $Sr_2IrO_4$ [43]. It would thus be interesting to investigate how the electron-boson coupling, pseudogap and d-wave instability respond to changes in the strength of the spin-orbit coupling.

In conclusion, although the origin of the boson mode in $Sr_2IrO_4$ and its interaction with other phenomena remain open questions, our study provides the first direct spectroscopic evidence for the existence of electron-boson coupling in the electron-doped $Sr_2IrO_4$. Combined with the earlier reports of pseudogap and d-wave instability, our study thus establishes $Sr_2IrO_4$ as a new platform to investigate the interplay between charge, spin and lattice degrees of freedom in doped spin-orbit coupled Mott insulators, which is believed to be a route towards many exotic phenomena.

We thank K.-J. Xu, S. Kumar and W. Mansuer for experimental help. We thank Y. Wang and Y. He for useful discussions. The work at the University of Science and Technology of China (USTC) was supported by the USTC start-up fund. The work at SLAC was supported by the U.S. DOE, Office of Basic Energy Science, Division of Materials Science and Engineering. SSRL is operated by the Office of Basic Energy Sciences, U.S. DOE, under Contract No. DE-AC02-76SF00515. The ARPES experiments at SSRL were carried out under User Proposal No. 4902 (PI J.-F. He). S.D.W. and X.C. acknowledge support from NSF Grant No. DMR-1905801. The work at Northeastern University was supported by the U.S. Department of Energy (DOE), Office of Science, Basic Energy Sciences Grant No. DE-FG02-07ER46352, and benefited from Northeastern University's Advanced Scientific Computation Center (ASCC) and the NERSC supercomputing center through DOE Grant No. DE-AC02-05CH11231.

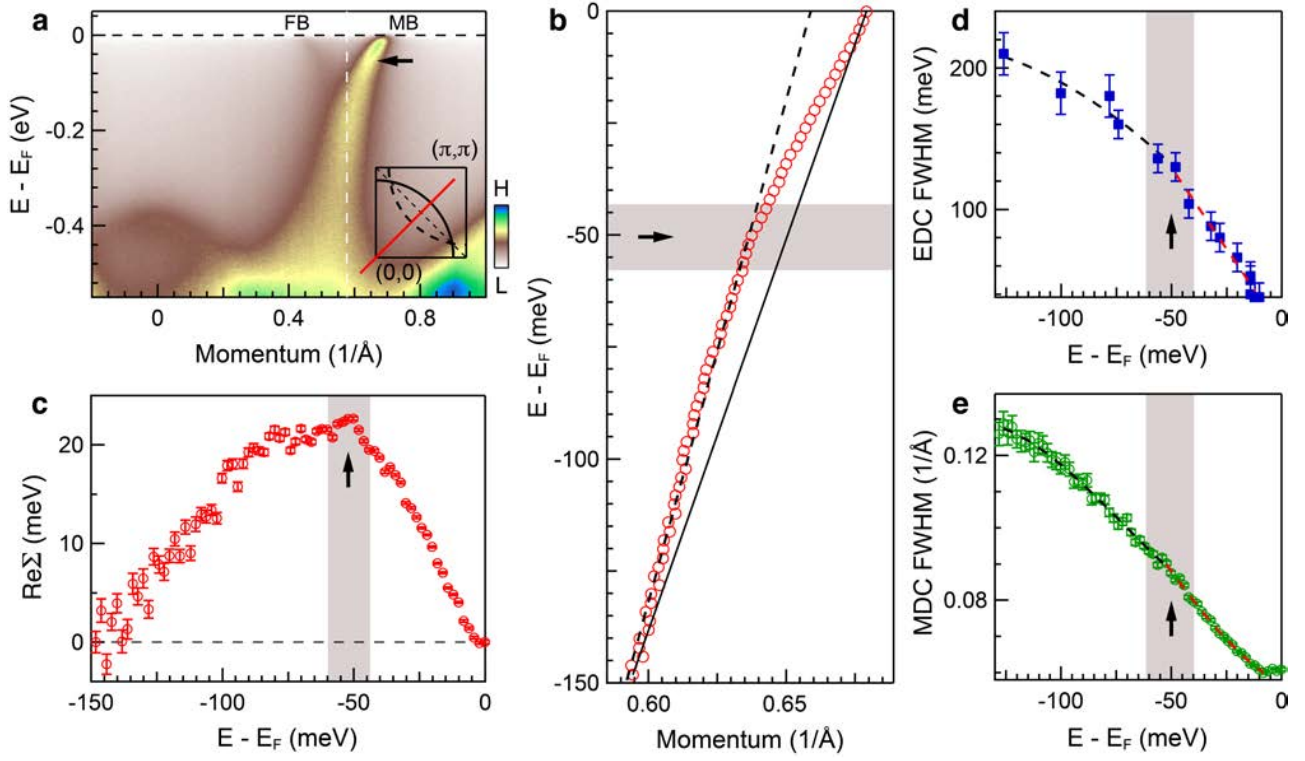

FIG. 1. Identification of the nodal kink in electron-doped $Sr_2IrO_4$. (a) Photoelectron intensity plot as a function of energy and momentum measured along the (0,0)-(π,π) nodal direction at 30K. The white-dashed line labels the reconstructed Brillouin zone boundary at (π/2, π/2). The black arrow marks the dispersion kink in the main band (MB). The energy feature in the folded band (FB, possibly associated with the octahedral rotation) is harder to identify due to its weak spectral intensity. (b) MDC-derived dispersion of the MB. The black straight line is the empirical bare band connecting the two energy points in the dispersion at $E_F$ and -150 meV. The dashed line is a guide for the eye. (c) Effective real part of electron self-energy. The black arrow marks the peak position at ~50 meV. Another possible weak feature locates at ~90 meV, which might be affected by its proximity to the FB. (d-e) Full-width-at-half-maximum (FWHM) of the EDC (d) and MDC (e) peaks. Black arrows label the drop in the peak width and dashed lines are a guide to the eye. The error bars represent the uncertainties in the determination of the peak width. Because of the doping limit in conventional chemical substitution, *in situ* potassium deposition was performed on a metallic $(Sr_{1-x}La_x)_2IrO_4$ sample (x~0.04) to reach a higher electron doping level [14, 21]. The estimated electron concentration n' is ~0.12 (see Fig. 4 and Supplemental Material for details).

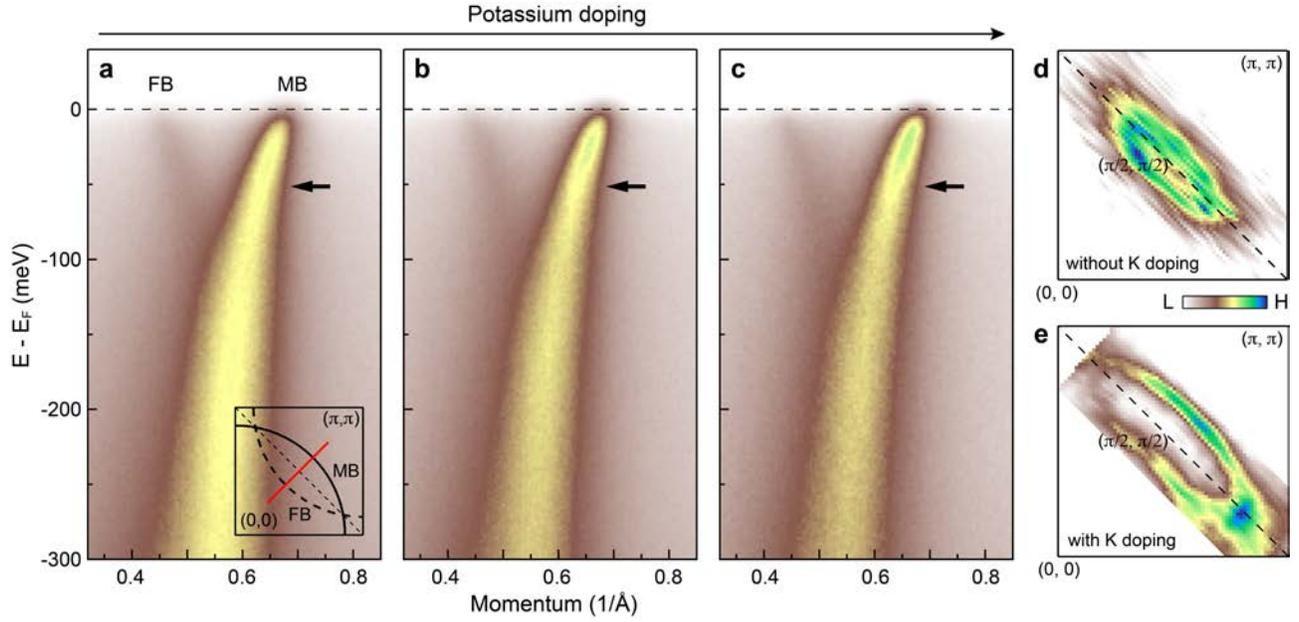

FIG. 2. Doping dependence of the electron band along the nodal direction. (a-c) Photoelectron intensity plots of the nodal dispersion as a function of doping at 30K. Continuous electron doping is achieved via *in situ* potassium deposition on a [$(Sr_{1-x}La_x)_2IrO_4$, x~0.04] sample. The black arrows indicate the kink position. The location of the momentum cut is shown in the inset of (a). (d) Fermi surface mapping of a [$(Sr_{1-x}La_x)_2IrO_4$, x~0.04] sample without potassium doping. (e) Same as (d), but for a [$(Sr_{1-x}La_x)_2IrO_4$, x~0.04] sample with potassium surface doping.

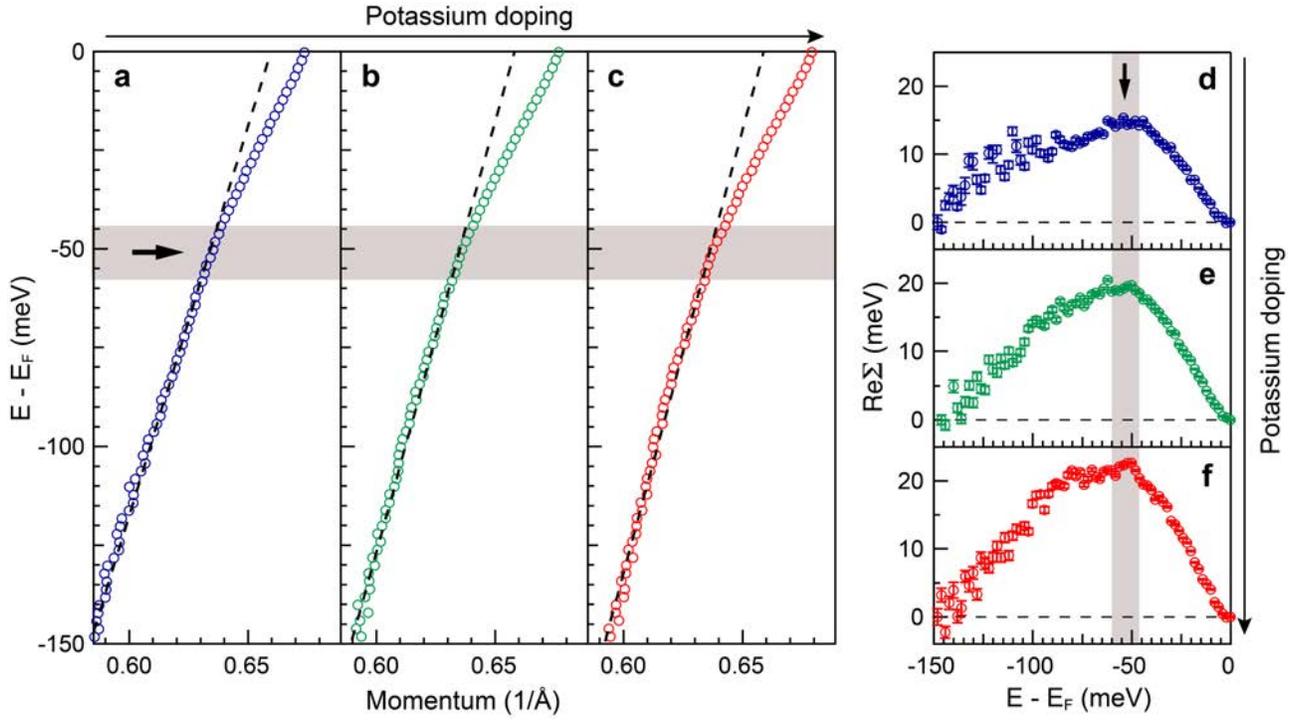

FIG. 3. Doping dependence of the nodal dispersion kink. (a-c) MDC-derived nodal dispersion extracted from the MB in Fig. 2(a-c), respectively. The black arrow marks the kink and the dashed lines are a guide to the eye. (d-f) Effective real parts of the electron self-energy obtained from the data in panels (a-c), respectively. The peak position as a function of doping is summarized in Fig. 4 as Sample 1 (blue circles).

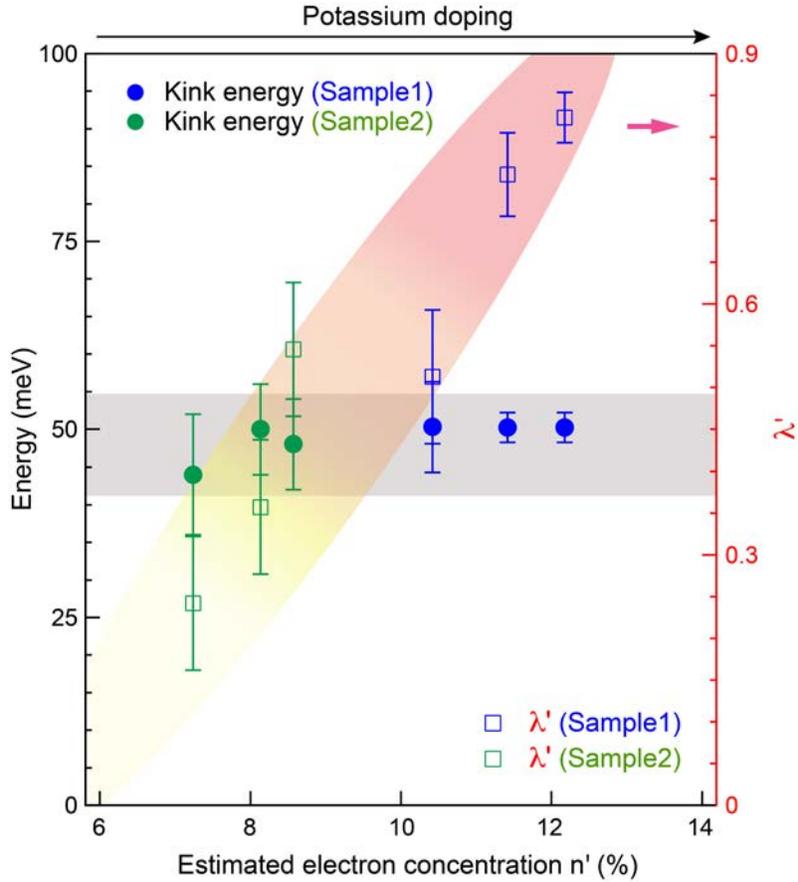

FIG. 4. Summary of the kink energy and band renormalization as a function of electron doping. In order to avoid excessive cycles of *in situ* potassium deposition, two samples were used to study the high and low doping regimes, respectively. Blue (green) circles represent the energy of the dispersion kink measured on Sample 1 (Sample 2). Empty squares indicate the renormalization factor λ' as defined in the main text. The estimated electron concentration n' is obtained by calculating the area of the underlying Fermi surface following the procedure used in [14]; this procedure underestimates electron concentration in the low doping regime, where the antinodal pseudogap exists [14] (also see Supplemental Material, Fig. S7). The error bars for the kink energy (renormalization factor λ') are from the uncertainties in the determination of the kink position (dispersion velocities).